\documentclass[twocolumn,amsmath,amssymb]{revtex4}
\usepackage{graphicx}
\usepackage{amssymb,amsmath}
\usepackage{dcolumn}
\usepackage{bm}

\begin{document}

\title{Coupled photonic crystal nanobeam cavities}
\author{Parag B. Deotare, Murray W. McCutcheon, Ian W. Frank, Mughees Khan, and Marko Lon\v{c}ar}
\affiliation{School of Engineering and Applied Sciences, Harvard University, Cambridge, MA 02138}

\date{\today}

\begin{abstract}

We describe the design, fabrication, and spectroscopy of coupled, 
high Quality ($Q$) factor silicon nanobeam photonic crystal cavities.
We show that the single nanobeam cavity modes are coupled 
into even and odd superposition modes, and we simulate the frequency and
$Q$ factor as a function of nanobeam spacing, demonstrating that a differential 
wavelength shift 
of 70 nm between the two modes is possible while maintaining $Q$ factors greater than 
$10^6$.  For both on-substrate and free-standing nanobeams, we experimentally monitor the 
response of the even mode as the gap is varied, and measure $Q$ factors as high as 
$2 \times 10^5$.

\end{abstract}

\maketitle

Photonic crystal cavities have been of great interest in recent years for their
ability to strongly confine light in wavelength-scale 
volumes~\cite{Noda05, Kuramochi06, Tanabe07}.  The high Quality ($Q$) factors 
and small mode volumes of these structures have enabled a wealth of new applications
in areas as diverse as low-threshold lasers~\cite{PainterScience,loncar_apl_2004}, optical 
switching~\cite{Tanabe05}, low power nonlinear optics~\cite{McCutcheonPRB}, cavity quantum 
electrodynamics~\cite{Hennessy07, Englund07}, and chemical sensing~\cite{Loncar, DiFalco,Kwon}.

Single photonic crystal (PhC) cavities have now been optimized to a point 
where they can be treated as fundamental photonic components.  To realize new 
functionality, it is essential to leverage the
natural scalability of PhC cavities.  For example, PhC cavities can 
be integrated into coupled resonator optical waveguides (CROWs)~\cite{Yariv_crow}
to realize novel heterostructures capable of slowing light~\cite{Notomi_Nphoton08}.
There is also much recent excitement about the possibility of
entangling the optical and mechanical degrees of 
freedom of a double cavity device\cite{Povinelli,Eichenfield_08, Chan_09}.  To achieve 
optomechanical coupling, the cavities require a small mass and a flexible 
platform.  These two properties are inherent to photonic crystal nanobeam 
cavities, which have been the subject of much recent investigation~\cite{Sauvan,Zain_08,
McCutcheon_08, Deotare09, Notomi_1D, Chan_09}.
In addition to optomechanical effects, double cavity structures could facilitate
adiabatic wavelength conversion.  It has been predicted that the spacing of 
double planar PhC cavities can be varied to realize broad bandwidth dynamic 
resonance tuning while maintaining the high $Q$ factor of the cavity mode~\cite{Notomi_PRL}.
Here, we show the static tuning of double cavity modes by varying the cavity separation, 
thereby demonstrating a proof-of-principle of this effect.

\begin{figure}[b]
\centering
\includegraphics[width=8.5cm]{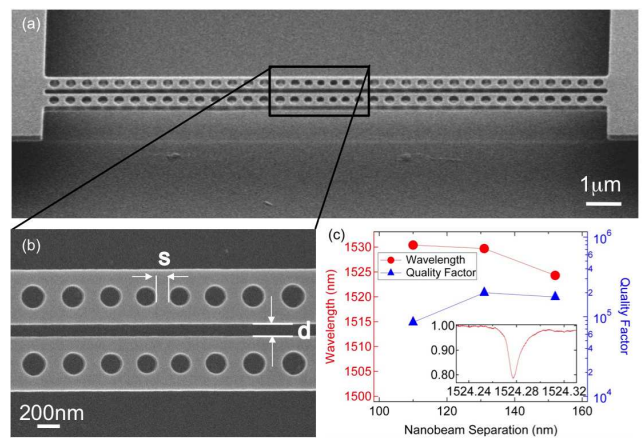}
\caption{(a,b) Double nanobeam cavity, showing the separation $d = 100$ nm, and cavity length,
$s = 146$ nm. (c) Mode wavelength and Quality factor for different nanobeam separations.  The
resonant scattering spectrum for the 150 nm double cavity is shown in the inset.
\label{fig:fab}}
\end{figure}

\begin{figure}[t]
\centering
\includegraphics[width=8.5cm]{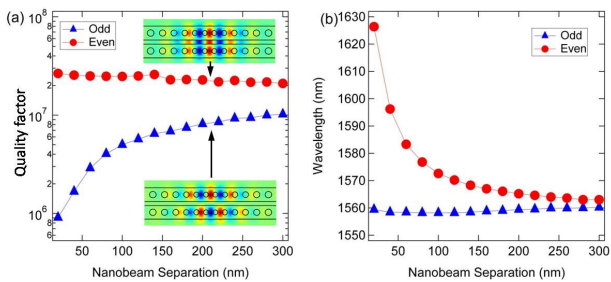}
\caption{(a) $Q$ factors of even and odd coupled nanobeam modes.  
Insets show $E_y$ components of the modes. (b) Mode wavelengths as a function of
separation, showing large dispersion of even mode.
\label{fig:sim}}
\end{figure}

In this work, we study coupled photonic crystal nanobeam cavities consisting
of two parallel suspended beams separated by a small gap, 
each patterned with a 1D line of holes, as shown in Fig.~\ref{fig:fab}(a).  
The starting point for our double nanobeam design is our previously reported single 
PhC nanobeam cavity~\cite{Deotare09, McCutcheon_08}.  To briefly summarize the approach, 
we start with a 220 nm thick and 500 nm wide free-standing Si 
waveguide (nanobeam) which supports only a single transverse electric mode.  The 
nanobeam is patterned with a linear array of air holes to introduce a stop-band
in the guided mode bandstructure around 1550 nm.  Near the middle of the beam, the hole
size and spacing are tapered to introduce a defect potential capable of strongly
localizing light.  Optimization of the adiabatic 5-hole taper design leads to 
simulated $Q$-factors greater than $10^7$, and we have experimentally 
measured $Q = 7.5 \times 10^5$ in these single beam structures~\cite{Deotare09}.
We then created double cavity structures by positioning two such cavities 
side-by-side and varying the air gap between them.  As expected from the 
physics of coupled harmonic oscillators, coupling generates new modes which are 
symmetric and anti-symmetric superpositions of the single cavity basis states, as
shown in Fig.~\ref{fig:sim}~\cite{Chan_09}. 
The energy splitting between the modes is dependent on the
strength of the coupling, which in this case is determined by the size of the gap.

The mode profiles, $Q$ factors, and resonant wavelengths are shown in Fig.~\ref{fig:sim}
as a function of the gap, $d$, between the nanobeams.  The data were simulated
with a 3D finite-difference time-domain code (Lumerical Solutions).
For small $d$, the coupling between the cavities is large, leading 
to a large splitting of the two modes.  The wavelength of the odd mode changes little
with $d$, whereas the even mode disperses rapidly to longer wavelength
as $d$ decreases.  This behavior can be understood from the mode profiles in 
Fig.~\ref{fig:sim}(a), which show that the even mode has an anti-node in the gap,
whereas the odd mode has a node in the gap.  The significant 
renormalization of the even mode renders it highly sensitive to $d$.  As $d$ shrinks 
below 100 nm, the field intensity of the gap anti-node
grows and becomes the dominant feature in the mode, much like an air-slot 
cavity~\cite{DiFalco,Foubert,Kwon,Almeida}.  At the same time, the $Q$ factor declines,
since the photonic crystal tapering is not optimized for the significant redistribution
of the mode energy into the slot between the nanobeams.  These factors imply that the even 
mode would be a sensitive probe of the inter-beam distance, and thus useful for 
optomechanical applications~\cite{Chan_09, Eichenfield_08}.  It would also be useful in 
bio- and chemical-sensing, as the gap anti-node would be highly sensitive to 
perturbations in the external environment.  As we will show below, it is only this mode 
which can be probed experimentally in our resonant scattering 
setup~\cite{McCutcheon_05,Deotare09}. 

In contrast to the significant alterations of the even mode, the odd mode is 
similar in profile to a superposition of two {\em uncoupled} 
single cavity modes.  Because of the node in the gap, 
the cavities do not significantly perturb each other, and this results in 
a relatively small dispersion with $d$.  The $Q$ factor, however, is 
substantially higher than that of the even mode, and in fact is approximately twice
the $Q$ factor of a single nanobeam cavity ($1.4 \times 10^7$).  Intuitively, this 
is consistent with the larger mode volume, $V \sim 0.7$($\lambda/n$)$^3$, which is about
twice that of the single cavity; the more extended real-space distribution of the fields 
results in a more localized $k$-space distribution, and therefore a reduction in the 
radiative components within the light cone~\cite{Srin_OE_03}.

The double nanobeam structures shown in Fig.~\ref{fig:fab} 
were fabricated on a silicon-on-insulator (SOI) wafer 
(SOITEC Inc) consisting of a silicon device layer of 220 nm, a SiO$_2$ layer of 
2$~\mu$m, and a thick silicon substrate.  A 
negative e-beam resist, Foxx-17 (Dow Corning) diluted in six parts of methyl isobutyl 
ketone, was spun onto the sample at 5000 rpm to 
give a layer 135 nm thick, and patterns were defined using a 100 kV electron beam 
lithography system (Elionix). A negative resist simplifies the pattern-writing, 
since the only exposed region is along the nanobeam,
and Foxx proved to be a robust etch mask for the RIE process. The resist was  
developed for 14 s in tetra-methyl ammonium hydroxide ($25$\% TMAH) followed by 
a thorough rinse in de-ionized water.  The patterns were transfered to the silicon 
layer using reactive ion etching with a SF$_6$, C$_4$F$_8$ and H$_2$ plasma.  The 
SiO$_2$ sacrificial layer was removed using a hydrofluoric acid vapor etching tool 
(AMMT)~\cite{deotare_hfvapor_2009,Deotare09}.  

We experimentally probed our double nanobeam cavities using a
cross-polarized resonant scattering technique~\cite{McCutcheon_05, Englund07}.
A tunable CW laser (Agilent) was focused onto the cavity from normal incidence with 
a microscope objective (numerical aperture = 0.5).  The resonantly scattered
reflected signal was analyzed in the cross polarization before being sent to an 
InGaAs detector.  Recently, we used the same approach to measure $Q$ factors as 
high as $7.5 \times 10^5$ in single nanobeam cavities~\cite{Deotare09}.  Because the 
resonant excitation field drives the ($E_y$) fields in the coupled nanobeams 
{\em in phase}, this technique is primarily sensitive to the even mode.  We note
that in principle, the field gradient of the focussed spot could be exploited to excite 
the odd mode, as could butt-coupling or evanescent waveguide coupling 
techniques~\cite{Srinivasan_04,Notomi04,Zain_08}.
Fig.~\ref{fig:fab}(c) shows the resonant wavelength and $Q$ factor of the even mode
for three different nanobeam separations.  As predicted by our simulations 
(Fig.~\ref{fig:sim}), the mode red-shifts as $d$ decreases.  
The $Q$ factor varies between $1 - 2 \times 10^5$.  Although this is more than an order
of magnitude lower than predicted by simulations, this is a highly useful range for
applications, and we expect increases as the fabrication quality of our structures
improves.

\begin{figure}[t]
\centering
\includegraphics[width = 8cm]{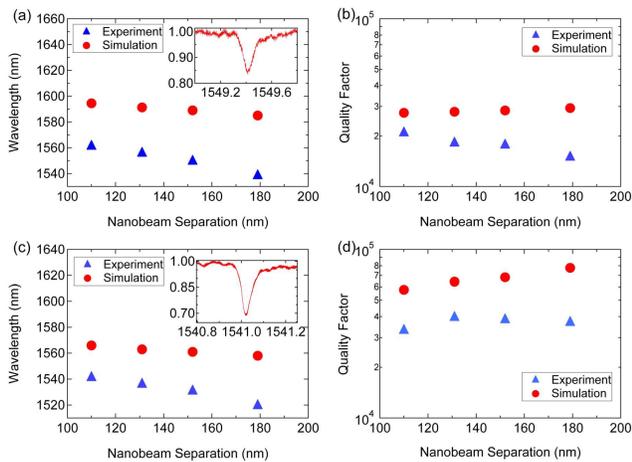}
\caption{Wavelength and $Q$ factor as a function of separation for 
on-substrate (a and b) and free-standing (c and d) double nanobeam cavities.
The insets show typical spectra normalized by reference spectra taken away from the 
cavity.  The simulations in (a) and (b) do not include the residual $\sim$ 40 nm 
Foxx resist layer, which has a negligible effect on the $Q$ factor and shifts the 
wavelength by only a few nm.
\label{fig:subrel}}
\end{figure}

We also investigated the effect of the substrate on the cavity modes.  As mentioned
above, for small $d$, the even mode field intensity is concentrated in the space
between the cavities.  This slot mode could be useful for bio- and chemical sensing, 
but to be robust in a liquid environment such a device would require the structural stability
provided by a substrate.  
Due to the limited tuning range of our laser, these data were obtained 
from cavities with a size ($s$) of 136 nm, which is smaller than our optimal design of 
146 nm (data shown in Fig.~\ref{fig:fab}(c)).
In Fig.~\ref{fig:subrel}(a) and (b), we present the results of our resonant scattering 
spectroscopy for unreleased, on-substrate nanobeam cavities (i.e. supported by SiO$_2$).   
The Foxx resist was not removed for these cavities, since any etch process would also 
remove the substrate.  However, we estimate that the resist layer remaining on top of the 
silicon nanobeams has a thickness less than 40 nm.  The resonant wavelength and $Q$ factor 
of the even mode are plotted as a function of $d$.  The
resonance blue-shifts with increasing $d$ as the effective index of the 
cavity mode decreases.
The experiment shows a larger dispersion than the simulation, likely reflecting the 
increasing role of e-beam lithography proximity effects for small gaps. 
The measured $Q$ factors of $1.5 - 2.0 \times 10^4$ are remarkably high for supported 
cavities~\cite{Zain_08}, 
considering that the designs were optimized for free-standing nanobeams.
Given the robust structure, high $Q$ factors, and field intensity in the gap, the 
supported double nanobeam cavity is a promising approach to scalable, on-chip
sensing applications.

Fig.~\ref{fig:subrel}(c) and (d) show the experimental data for the same cavities 
after the sacrifical
SiO$_2$ substrate was removed.  The modes are blue-shifted compared to the results
of Fig.~\ref{fig:subrel}(a), as expected from the decreased effective index of the air cladding
compared to the oxide.  The $Q$ factors, however, are increased by a factor of $\sim 2$ due 
to the reduced leakage into the substrate.   

In conclusion, by varying the spacing of two coupled nanobeam cavities, 
we have achieved resonance 
tuning over a 20 nm range for both on-substrate and free-standing structures while 
maintaining a relatively constant $Q$ factor.  This shows the same tuning principle that
was predicted recently for double planar photonic crystal 
cavities~\cite{Notomi_PRL}, and the effect would be magnified for coupled
nanobeams with smaller separations.   The ability to tune the resonant wavelength of a cavity 
without adverse effects on the $Q$ factor opens up new potential areas 
for research in optomechanics, adiabatic frequency conversion, and sensing. 

\section*{Acknowledgements}
This work is supported in part by Harvard's National Science and Engineering Center, 
NSF grant ECCS-0701417, and NSF CAREER grant. Device fabrication was performed at 
the Center for Nanoscale Systems at Harvard. MWM would like to thank NSERC (Canada) for 
its support, and IWF thanks the NSF GRFP.

\bibliographystyle{unsrt}

\begin{thebibliography}{10}

\bibitem{Noda05}
B.~S. Song, S.~Noda, T.~Asano, and Y.~Akahane.
\newblock Ultra-high-{$Q$} photonic double-heterostructure nanocavity.
\newblock {\em Nature Materials}, 4:207--210, 2005.

\bibitem{Kuramochi06}
E.~Kuramochi, M.~Notomi, S.~Mitsugi, A.~Shinya, T.~Tanabe, and T.~Watanabe.
\newblock Ultrahigh-{$Q$} photonic crystal nanocavities realized by the local
  width modulation of a line defect.
\newblock {\em Appl.\ Phys.\ Lett.}, 88:041112, 2006.

\bibitem{Tanabe07}
T.~Tanabe, M.~Notomi, E.~Kuramochi, A.~Shinya, and H.~Taniyama.
\newblock Trapping and delaying photons for one nanosecond in an ultrasmall
  high-q photonic-crystal nanocavity.
\newblock {\em Nature Photonics}, 1:49--52, 2007.

\bibitem{PainterScience}
O.~Painter, R.~K. Lee, A.~Scherer, A.~Yariv, J.~D. O'Brien, P.~D. Dapkus, and
  I.~Kim.
\newblock Two-dimensional photonic band-gap defect mode laser.
\newblock {\em Science}, 284:1819--1821, 1999.

\bibitem{loncar_apl_2004}
M.~Loncar, T.~Yoshie, A.~Scherer, P.~Gogna, and Y.~M. Qiu.
\newblock Low-threshold photonic crystal laser.
\newblock {\em Appl.\ Phys.\ Lett.}, 81:2680--2682, 2002.

\bibitem{Tanabe05}
T.~Tanabe, M.~Notomi, S.~Mitsugi, A.~Shinya, and E.~Kuramochi.
\newblock All-optical switches on a silicon chip realized using photonic
  crystal nanocavities.
\newblock {\em Appl.\ Phys.\ Lett.}, 87:151112, 2005.

\bibitem{McCutcheonPRB}
M.~W. McCutcheon, J.~F. Young, G.~W. Rieger, D.~Dalacu, S.~Fr\'{e}d\'{e}rick,
  P.~J. Poole, and R.~L. Williams.
\newblock Experimental demonstration of second-order processes in photonic
  crystal microcavities at submilliwatt excitation powers.
\newblock {\em Phys.\ Rev.\ B}, 76:245104, 2007.

\bibitem{Hennessy07}
K.~Hennessy, A.~Badolato, M.~Winger, D.~Gerace, M.~Atature, S.~Gulde, S.~Falt,
  E.~L. Hu, and A.~Imamoglu.
\newblock Quantum nature of a strongly coupled single quantum dot-cavity
  system.
\newblock {\em Nature}, 445:896--899, 2007.

\bibitem{Englund07}
D.~Englund, A.~Faraon, I.~Fushman, N.~Stoltz, P.~Petroff, and
  J.~Vu\v{c}kovi\'{c}.
\newblock Controlling cavity reflectivity with a single quantum dot.
\newblock {\em Nature}, 450:857--861, 2007.

\bibitem{Loncar}
M.~Lon\v{c}ar, A~. Scherer, and Y.~Qiu.
\newblock Photonic crystal laser sources for chemical detection.
\newblock {\em Appl.\ Phys.\ Lett.}, 82:4648--4650, 2003.

\bibitem{DiFalco}
A.~Di Falco, L.~O'Faolain, and T.~F. Krauss.
\newblock Chemical sensing in slotted photonic crystal heterostructure
  cavities.
\newblock {\em Appl.\ Phys.\ Lett.}, 94:063503, 2009.

\bibitem{Kwon}
S.-H. Kwon, T.~Sunner, M.~Kamp, and A.~Forchel.
\newblock Optimization of photonic crystal cavity for chemical sensing.
\newblock {\em Opt. Express}, 16:11709--11717, 2008.

\bibitem{Yariv_crow}
A.~Yariv, Y.~Xu, R.~K. Lee, and A.~Scherer.
\newblock Coupled-resonator optical waveguide: a proposal and analysis.
\newblock {\em Opt. Lett.}, 24:711713, 1999.

\bibitem{Notomi_Nphoton08}
M.~Notomi, E.~Kuramochi, and T.~Tanabe.
\newblock Large-scale arrays of ultrahigh-$q$ coupled nanocavities.
\newblock {\em Nature Photonics}, 2:741--747, 2008.

\bibitem{Povinelli}
M.~L. Povinelli, M.~Loncar, M.~Ibanescu, E.~J. Smythe, S.~G. Johnson,
  F.~Capasso, and J.~D. Joannopoulos.
\newblock Evanescent-wave bonding between optical waveguides.
\newblock {\em Opt. Lett.}, 30:3042--3044, 2005.

\bibitem{Eichenfield_08}
M.~Eichenfield, R.~Camacho, J.~Chan, K.~J. Vahala, and O.~Painter.
\newblock A picogram and nanometer scale photonic crystal opto-mechanical
  cavity.
\newblock {\em arXiv:0812.2953v1 [physics.optics]}, 2008.

\bibitem{Chan_09}
J.~Chan, M.~Eichenfield, R.~Camacho, and O.~Painter.
\newblock Optical and mechanical design of a ''zipper'' photonic crystal
  optomechanical cavity.
\newblock {\em Opt. Express}, 17:3802--3817, 2009.

\bibitem{Sauvan}
C.~Sauvan, G.~Lecamp, P.~Lalanne, and J.P. Hugonin.
\newblock Modal-reflectivity enhancement by geometry tuning in photonic crystal
  microcavities.
\newblock {\em Opt. Express}, 13:245--255, 2005.

\bibitem{Zain_08}
A.~R.~M. Zain, N.~P. Johnson, M.~Sorel, and R.~M.~De la~Rue.
\newblock Ultra high {Q}uality factor one dimensional photonic crystal/photonic
  wire micro-cavities in silicon-on-insulator {(SOI)}.
\newblock {\em Opt. Express}, 16:12084--12089, 2008.

\bibitem{McCutcheon_08}
M.~W. McCutcheon and M.~Lon\v{c}ar.
\newblock Design of a silicon nitride photonic crystal nanocavity with a
  {Q}uality factor of one million for coupling to a diamond nanocrystal.
\newblock {\em Opt. Express}, 16:19136--19145, 2008.

\bibitem{Deotare09}
P.~B. Deotare, M.~W. McCutcheon, I.~W. Frank, M.~Khan, and M.~Loncar.
\newblock High quality factor photonic crystal nanobeam cavities.
\newblock {\em Appl.\ Phys.\ Lett.}, 94:121106, 2009.

\bibitem{Notomi_1D}
M.~Notomi, E.~Kuramochi, and H.~Taniyama.
\newblock Ultrahigh-${Q}$ nanocavity with 1{D} photonic gap.
\newblock {\em Opt. Express}, 16:11095--11102, 2008.

\bibitem{Notomi_PRL}
M.~Notomi, H.~Taniyama, S.~Mitsugi, and E.~Kuramochi.
\newblock Optomechanical wavelength and energy conversion in high-$q$
  double-layer cavities of photonic crystal slabs.
\newblock {\em Phys. Rev. Lett.}, 97, 2006.

\bibitem{Foubert}
K.~Foubert, L.~Lalouat, B.~Cluzel, E.~Picard, D.~Peyrade, E.~Delamadeleine,
  F.~de~Fornel, and E.~Hadji.
\newblock Near-field modal microscopy of subwavelength light confinement in
  multimode silicon slot waveguides.
\newblock {\em Appl.\ Phys.\ Lett.}, 93:251103, 2008.

\bibitem{Almeida}
V.~R. Almeida, Q.~Xu, C.~A. Barrios, and M.~Lipson.
\newblock Guiding and confining light in void nanostructure.
\newblock {\em Opt.\ Lett.}, 29:1209--1211, 2004.

\bibitem{McCutcheon_05}
M.~W. McCutcheon, G.~W. Rieger, I.~W. Cheung, J.~F. Young, D.~Dalacu,
  S.~Fr{\'e}d{\'e}rick, P.~J. Poole, G.~C. Aers, and R.~L. Williams.
\newblock Resonant scattering and second-harmonic spectroscopy of planar
  photonic crystal microcavities.
\newblock {\em Appl.\ Phys.\ Lett.}, 87:221110, 2005.

\bibitem{Srin_OE_03}
K.~Srinivasan and O.~Painter.
\newblock Momentum space design of high-{$Q$} photonic crystal optical
  cavities.
\newblock {\em Opt. Express}, 10:670--684, 2003.

\bibitem{deotare_hfvapor_2009}
P.~B. Deotare, M.~Khan, and M.~Loncar.
\newblock Vapor phase release of silicon nanostructures for optomechanics
  applications.
\newblock {\em Proc. SPIE}, 7205:7205--09, 2009.

\bibitem{Srinivasan_04}
K.~Srinivasan, P.~E. Barclay, M.~Borselli, and O.~Painter.
\newblock Optical-fiber-based measurement of an ultrasmall volume high-{$Q$}
  photonic crystal microcavity.
\newblock {\em Phys.\ Rev.\ B}, 70:081306, 2004.

\bibitem{Notomi04}
M.~Notomi, A.~Shinya, S.~Mitsugi, E.~Kuramochi, and H-Y. Ryu.
\newblock Waveguides, resonators and their coupled elements in photonic crystal
  slabs.
\newblock {\em Opt. Express}, 12:1551--1561, 2004.

\end{thebibliography}

\end{document}